\documentclass[preprint,12pt]{elsarticle}

\usepackage{graphicx} 
\usepackage{amsmath} 
\usepackage{flushend} 
\usepackage[ruled,linesnumbered]{algorithm2e} 
\usepackage{amssymb}
\usepackage{booktabs}
\usepackage{float}
\usepackage{array}
\usepackage{subfigure}
\usepackage{multirow} 
\usepackage{tabularx} 
\usepackage{makecell}  
\journal{Nuclear Physics B}

\begin{document}

\begin{frontmatter}



\title{Multi-Perspective Attention Mechanism for Bias-Aware Sequential Recommendation}


\author[label1]{Mingjian Fu}\ead{sinceway@fzu.edu.cn} 
\author[label1]{Hengsheng Chen}\ead{231027050@fzu.edu.cn} 
\author[label1]{Dongchun Jiang}\ead{948234053@qq.com} 
\author[label1]{Yanchao Tan\corref{cor}}\ead{yctan@fzu.edu.cn}
\cortext[cor]{Corresponding author.}
\affiliation[label1]{organization={College of Computer and Data Science},
            addressline={Fuzhou University}, 
            city={Fuzhou},
            postcode={350108}, 
            country={China}}
        
\begin{abstract}
In the era of advancing information technology, recommender systems have emerged as crucial tools for dealing with information overload. However, traditional recommender systems still have limitations in capturing the dynamic evolution of user behavior. To better understand and predict user behavior, especially taking into account the complexity of temporal evolution, sequential recommender systems have gradually become the focus of research. Currently, many sequential recommendation algorithms ignore the amplification effects of prevalent biases, which leads to recommendation results being susceptible to the Matthew Effect. Additionally, it will impose limitations on the recommender system's ability to deeply perceive and capture the dynamic shifts in user preferences, thereby diminishing the extent of its recommendation reach. To address this issue effectively, we propose a recommendation system based on sequential information and attention mechanism called Multi-Perspective Attention Bias Sequential Recommendation (MABSRec). Firstly, we reconstruct user sequences into three short types and utilize graph neural networks for item weighting. Subsequently, an adaptive multi-bias perspective attention module is proposed to enhance the accuracy of recommendations. Experimental results show that the MABSRec model exhibits significant advantages in all evaluation metrics, demonstrating its excellent performance in the sequence recommendation task. 
\end{abstract}




\end{frontmatter}



\section{Introduction}
The rapid advancement of information technology has led to significant progress in the production, storage, dissemination, and acquisition of information. However, this trend has also resulted in an unprecedented explosion of information that far exceeds the processing capabilities of individuals and organizations, thereby diminishing the utility of information \cite{10.1111/jcc4.12178}, \cite{doi:10.1287/orsc.1100.0634}. Meanwhile, this growth in information also poses serious challenges to the design and management of information systems \cite{10.3389/fpsyg.2023.1122200}. In this context, recommender systems have emerged as a powerful tool to address the challenge of information overload effectively. These systems provide highly personalized recommendations based on users' historical behavior and interests. Moreover, by leveraging implicit user feedback, such as clicks and purchases, recommender systems can perform deep learning and deliver personalized recommendations without relying solely on explicit search keywords \cite{4781121}. Therefore, in the context of information overload, recommendation systems have significantly optimized the experience of information retrieval, providing users with more personalized and diversified information services.

The application of recommender systems is extensive, spanning various fields such as e-commerce \cite{10.1145/3411564.3411572}, social media \cite{10.1145/1458082.1458205}, and tourism \cite{Braunhofer2014TechniquesFC}. Traditional recommender systems are generally classified into three main categories: collaborative filtering-based \cite{10.1145/3460231.3478854}, \cite{FANG2022109044}, \cite{10.1145/3459637.3482354}, content-based \cite{HUANG2022108596}, \cite{DELCARMENRODRIGUEZHERNANDEZ2021106740}, \cite{9773925}, and hybrid recommender systems \cite{10.1145/3460231.3474272}, \cite{10.1145/3511808.3557354}, \cite{9723533}. Collaborative filtering-based recommender systems utilize cosine similarity, Pearson's correlation coefficient, and other computational methods to measure the similarity between users or items to recommend items of similar interest to users. Content-based recommender systems focus on analyzing the characteristic attributes of items to achieve personalized recommendations by gaining insight into the user's interest in these characteristics. Hybrid recommender systems combine collaborative filtering-based and content-based recommender systems, aiming to overcome the shortcomings of a single recommendation algorithm and provide more accurate and diverse personalized recommendations. Although traditional systems have achieved notable success in meeting user needs, they struggle to capture the dynamic evolution of user behavior over time. To this end, the researchers carried out a study on sequence recommendation. Sequential recommender systems have been developed to emphasize temporal information, enabling better understanding and predicting user behavior, particularly its temporal evolution \cite{10.1145/3511808.3557268}. Bao et al. \cite{10.1145/3604915.3608857} proposed TALLRec to efficiently train Large Language Models (LLMs) by transforming recommendation data into instructions using two fine-tuning techniques, Alpaca Tuning and Rec-Tuning. Ren et al. \cite{ijcai2023p254} proposed a review-based recommendation method that leverages a self-supervised graph decomposition network. This network learns separate representations of users and items on latent factors through a graph decomposition learning module. Additionally, they introduced an intent-aware contrastive learning task to alleviate data sparsity and enhance the separation of user and item representations. Du et al. \cite{10.1145/3539618.3591679} proposed a framework called EMKD to improve the accuracy of sequence recommendation by training multiple parallel networks and performing comparative knowledge distillation. 

\begin{figure*}[h]
    \centering
    \includegraphics[width=1\linewidth]{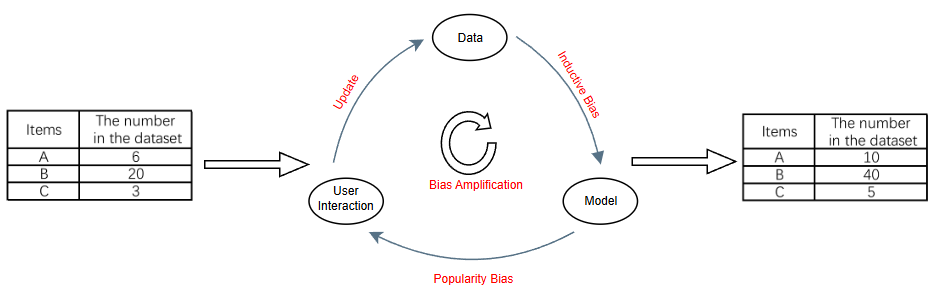}
    \caption{Matthew Effect in Recommendation}
    \label{fig:Matthew}
\end{figure*}

Although existing research on sequential recommender systems has demonstrated commendable performance, studies on the impact of various bias factors present in user data remain insufficient. Notably, prevalent biases such as popularity bias and amplified subjective bias are frequently observed in user-item interactions \cite{Guo2024}. Failure to account for these prevalent biases in recommender systems often leads to the manifestation of the Matthew Effect \cite{9778074}. To solve this problem, we propose a novel sequence recommendation model, called Multi-Perspective Attention Bias Sequential Recommendation (MABSRec). The model constructs multiple bias views from multiple bias perspectives, explicitly incorporating both popular bias and amplified subjective bias in user data. Subsequently, a graph convolution operation is used to enrich the representation of each item within each bias view. Finally, an attention fusion network is utilized to weigh the impact of various biases on users and output the predicted results. Experimental evaluations conducted on three real-world datasets indicate that the MABSRec model has significant advantages in all evaluation indexes, showing its excellent performance in the sequence recommendation task. The principal contributions of this study are delineated as follows:
\begin{itemize}
    \item We consider the prevalent bias in the user data and the amplified subjective bias, and divide the user sequence data into three biased short sequences: popularity-biased short sequences, subjectivity-biased short sequences, and debiased short sequences. Then, by constructing a graph convolutional neural network, the three kinds of biased sequences are constructed into the corresponding sequence graph information, so that the sequence recommendation model can more deeply mine the biased information of other users. 
    \item We adopt the multi-head self-attention mechanism with shared weights to capture the key information in the user sequence to capture the dynamic interest changes of users better. In addition, we introduce an adaptive multi-perspective attention bias module to fuse the three bias information to comprehensively consider the impact of different biases on user behavior.
    \item We comprehensively evaluate the proposed MABSRec model across multiple dimensions using three real-world datasets. Extensive experimental results substantiate the effectiveness and superiority of MABSRec. Additionally, ablation studies confirm the rationality and efficacy of the key components.
\end{itemize} 

The rest of this article is structured as follows: Section \ref{related work} provides a review of related work. Section \ref{method} introduces the problem definition and details of the proposed MABSRec model. Section \ref{experimental} evaluates the performance of the proposed MABSRec model and verifies the effectiveness of key components. Section \ref{conclusion} summarizes this work.

\section{Related Work}
\label{related work}
Sequence recommendation is one of the crucial tasks in recommender systems, and its core goal is to accurately predict the possible future sequence of items of interest based on the user's historical behavior sequence. At the early stage of the research on sequence recommendation, scholars drew on methods such as Markov chains to deal with the recommendation problem of sequence information \cite{10.1145/1772690.1772773}, \cite{10.1145/2766462.2767694}. However, with the rapid development of deep neural networks, deep learning models such as Convolutional Neural Network (CNN) and Recurrent Neural Network (RNN) have been introduced to sequence recommendation tasks. Although these traditional methods have achieved some success, they generally suffer from the long-tail effect, slow training speed, and limited ability to extract temporal features in long sequence recommendation scenarios.

In recent years, the attention mechanism, Graph Neural Network (GNN), and Contrastive Learning (CL) have been applied to sequence recommendation models due to their powerful comprehension capabilities, and have become mainstream research directions. In this section, we review the sequence recommendation models using these techniques.

\subsection{Attention Mechanism}
The attentional mechanism, an approach inspired by the human visual and cognitive systems, allows neural networks to selectively focus on important information within the input data during processing. This mechanism effectively improves the performance and generalization of the model. Zhou et al. \cite{10.1609/aaai.v33i01.33015941} introduced the attention mechanism into the Gated Recurrent Unit (GRU) model and designed the AUGRU module to cope with the problem of user interest drift. This method enables the model to better adapt to changes in user interests, however, it significantly reduces the computational efficiency of the network due to the high complexity of the GRU module in the computation process. Kang et al. \cite{8594844} proposed the SASRec recommendation model, an adaptation of the Transformer architecture \cite{10.5555/3295222.3295349}. SASRec introduces trainable positional embeddings, enabling the model to differentiate items based on their positional context. Additionally, it employs a multi-head self-attention mechanism to capture intricate relationships within sequences, thereby enhancing the model's ability to represent inter-item dependencies. Shin et al. \cite{Shin_Choi_Wi_Park_2024} proposed the BSARec model to balances the strengths of both approaches and mitigates the over-smoothing problem, which combines the Fourier transform and the self-attention mechanism. Liu et al. \cite{10.1145/3539618.3591717} proposed a novel linear attention mechanism in long sequence recommendation systems, named LinRec. LinRec reduces the complexity of the Transformer by changing the dot product order, $L_2$-normalization, and ELU activation, while maintaining the accuracy.

\subsection{Graph Neural Network}
GNN is a class of deep learning models specifically designed for processing graph-structured data, among which the Graph Convolutional Network (GCN) \cite{kipf2017semisupervised} is one of the most representative models. GCN extracts feature information by performing convolutional operations on graphs to achieve deep learning and analysis of graph data. In recent years, many researchers have applied graph neural networks to the recommendation domain, achieving significant experimental results. Wang et al. proposed \cite{10.1145/3331184.3331267} an innovative recommendation method, called NGCF, which considers both users and items as nodes in a graph structure and constructs an information network using user-item interaction records. By representing users and items as nodes and capturing their interactions, NGCF effectively extends user and item representations. LightGCN \cite{10.1145/3397271.3401063} simplifies the model structure by using a simple aggregation weighting method to enhance the efficiency of model training and implementation. Specifically, LightGCN utilizes the user-item interaction matrix to construct the graph structure and updates node features through weighted aggregation. Chang et al. \cite{10.1145/3404835.3462968} proposed a metric learning-based approach for building user interest graph structures. This method first constructs a dynamic graph by measuring node similarity as a key metric and connecting two nodes only when their similarity exceeds a predefined threshold. Subsequently, the importance of nodes surrounding the target node is modeled by computing attention scores between the target node and its neighbors, as well as the attention coefficients of neighboring nodes relative to the target node. TransGNN \cite{10.1145/3626772.3657721} combines the strengths of the Transformer and GNN to help the GNN expand its receptive field. TransGNN uses three types of positional encodings to capture graph structural information and then alternates between the Transformer and GNN layers to focus each node on the most relevant samples.

\subsection{Contrastive Learning}
The core goal of CL is to achieve effective data representation learning by minimizing the distance between positive samples and a given anchor while maximizing the distance between negative samples and the same anchor. In sequence recommendation, CL can address the challenges posed by sparse interaction data and enhance the learning of effective representations \cite{lee2023hierarchicalcontrastivelearningmultiple}, \cite{10.1145/3539618.3591692}. CL4SRec \cite{9835621} is the first to introduce CL into the sequential recommendation domain. CL4SRec combines the traditional sequential prediction objective with a CL objective to improve the accuracy and versatility of recommender systems. Specifically, CL4SRec constructs user sequences from different perspectives and utilizes a contrastive loss function to learn more accurate user representations. DuoRec \cite{10.1145/3488560.3498433} improves the distribution of sequence representations and item embeddings by introducing CL regularization while utilizing both unsupervised and supervised contrastive samples. DCRec \cite{10.1145/3543507.3583361} improves the accuracy and diversity of recommender systems by unifying sequential pattern encoding with global synergistic relationship modeling through adaptive consistency-aware augmentation. In addition, DCRec utilizes CL for self-supervised signal extraction across different views, effectively capturing intra-sequence item transition patterns and inter-sequence user dependencies.

\section{The Proposed Method}
\label{method}
\begin{figure*}
    \centering
    \includegraphics[width=1\linewidth]{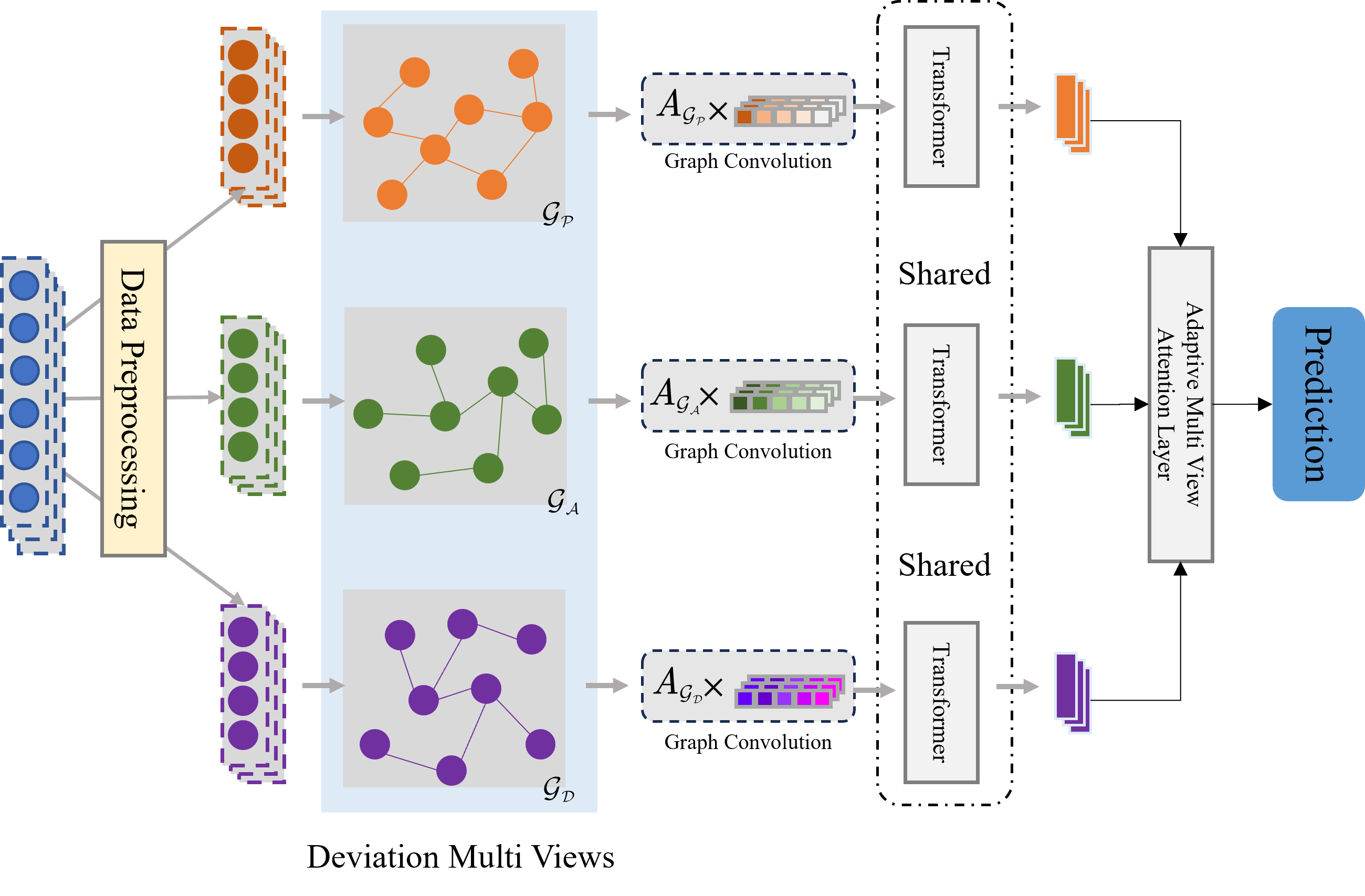}
    \caption{Framework diagram of the MABSRec model}
    \label{fig: framework}
\end{figure*}

In this section, we first define the sequence recommendation problem. Then, we illustrate the novel framework MABSRec
in detail. The framework is shown in Figure \ref{fig: framework}. Finally, we present the definition of the loss function.

\subsection{Problem Definition}
For the sequence recommendation problem, let $U$ and $I$ denote a set of users and a set of items, respectively. For each user $u \in U$, their historical interaction sequence can be defined as $S_u=\left(s_u^1, s_u^2, \cdots, s_u^t\right)$, where $s_u^j$ denotes the user's interaction with item $i \in I$ at the $j$ time step. $t$ denotes the length of the historical interaction sequences of different users, and $j$ denotes the position of item $i$ in the historical interaction sequence $S_u$ of user $u$. The goal of sequential recommendation is to maximize the likelihood of inferring from a collection of items the items that a user is most likely to interact with next, by giving a history of interactions. The objective function is as follows:
\begin{equation}
\arg \max _{i \in I} P\left(s_u^{t+1}=i \mid S_u\right)
\end{equation}

Since different users may have different sequence lengths, to maintain consistency, we perform complementary zero padding for sequence lengths less than the preset length and intercept sequence lengths that are more than the preset length.

\subsection{Multi-bias Processing in Data}
\begin{figure*}[ht]
    \centering
    \includegraphics[width=1\linewidth]{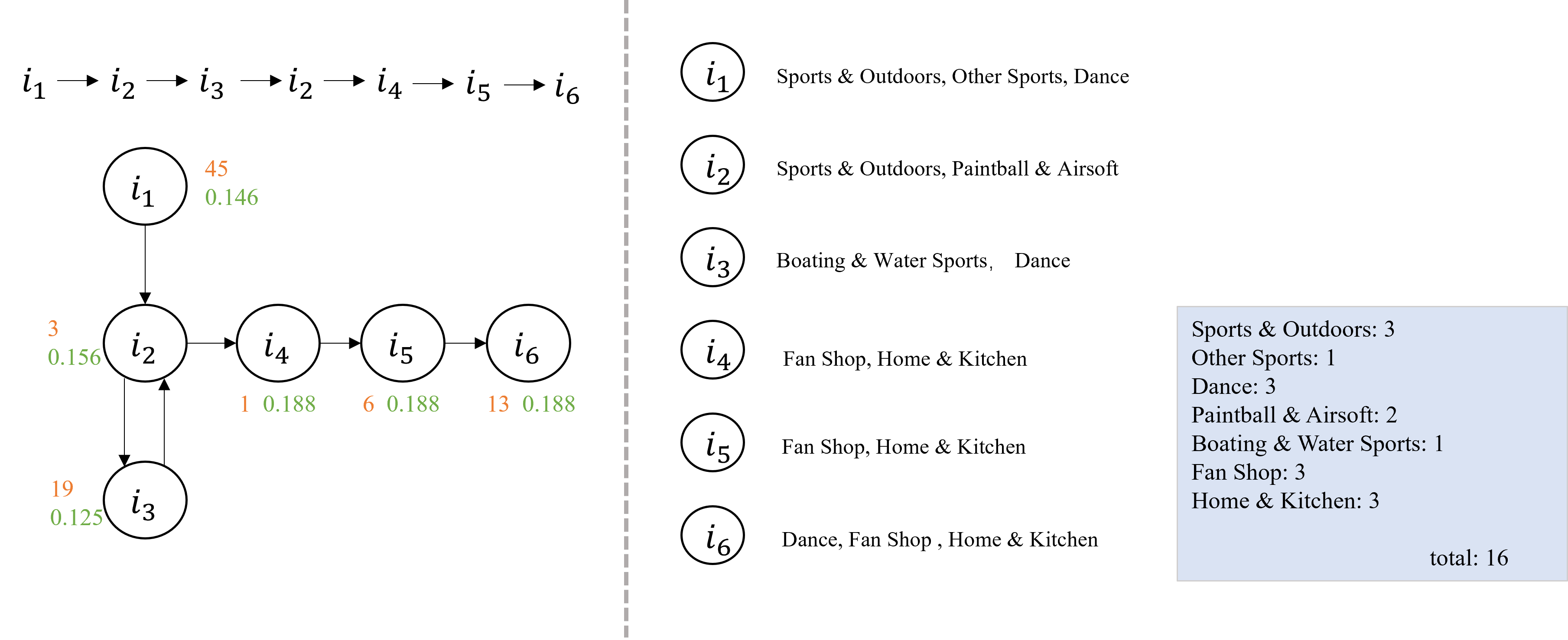}
    \caption{Calculation of the degree of bias}
    \label{fig: calculation}
\end{figure*}

In recommender systems, data is often subject to popularity bias and amplified subjective bias. To visualize the popularity that each item receives in the dataset, MABSRec uses the number of times an item appears in the dataset to represent its popularity, denoted as $S_p$. The more times an item appears, the more popular it is. In addition, for the personalized preferences in the user sequences, the user's personalized weights for the items are defined as $S_c$. The weights of the items are calculated as follows:

\begin{equation}
\left\{\begin{array}{l}
S_{p}(u, i) = count(i) \\
S_{c}(u, i) = \frac{c_{i} \cdot c_{u}}{\left|c_{i}\right|}
\end{array}\right.
\end{equation}

where $count(i)$ denotes the count of the number of occurrences of each item $i$ in the dataset, $c_i$ is the one-hot representation of the category of item $i$ in a sequence. $\left|c_i\right|$ denotes the number of categories corresponding to the item. $c_u$ denotes the vector of the number of occurrences of all the categories in the sequence. By considering $S_p$ and $S_c$, MABSRec can more accurately understand the prevalence bias and the user's personalized preference in the recommender system. As shown in Figure \ref{fig: calculation}, 
all items in the sequence have seven category attributes: "Sports $\&$ Outdoors", "Other Sports", "Dance", "Paintball $\&$ Airsoft", "Boating $\&$ Water Sports", "Fan Shop", and "Home $\&$ Kitchen". For item $i_{1}$, it contains three categories: "Sports $\&$ Outdoors", "Other Sports", and "Dance". Therefore, the category one-hot corresponding to the item $i_{1}$ is denoted as $[1,1,1,0,0,0,0]$, and $\left|c_{i}\right|$ is 3, which denotes the number of categories of item $i_{1}$. $c_{u}$ is the vector of the number of occurrences of all the sequence categories, denoted as $[3,1,3,2,1,3,3]$. In addition, the orange numbers indicate the number of times each item appears in the dataset, i.e., $S_{p}$, and the green numbers indicate the weight, i.e., $S_{c}$, computed based on the category or other attributes of each item in the sequence.

Then, we filter and reorganize the items in the user sequences from three biased perspectives. We categorize those items that are popular among the entire user community but account for less in terms of personalized preferences as popular bias perspective items, denoted as $I_\mathcal{P}^\mathcal{S}$, and computed by the following formula:
\begin{align}
   I_{\mathcal{A}}^{S}=&\left\{i\mid rank\left(S_{_p}(u,i)>k_{\mathcal{P}}\cdot|I_{\mathcal{S}}|\right), \notag\right. \\ &\left. rank\left(S_{_c}(u,i)<k_{\mathcal{A}}\cdot|I_{\mathcal{S}}|\right),i\in I_{\mathcal{S}}\right\}
\end{align}
where, $I_{\mathcal{S}}$ denotes the set of all items in an interaction sequence, and $k_{\mathcal{P}}$, $k_{\mathcal{A}}$ are the degree of popularity bias and amplified subjective bias, respectively. 

In addition, we take those items with low audience degree but in line with the user's personality as amplified subjective bias perspective items, denoted as $I_{\mathcal{A}}^\mathcal{S}$, which are calculated as follows:
\begin{align}
    I_{\mathcal{A}}^{S}=&\left\{i\mid rank\left(S_{_p}(u,i)>k_{\mathcal{P}}\cdot|I_{\mathcal{S}}|\right), \notag\right. \\ &\left. rank\left(S_{_c}(u,i)<k_{\mathcal{A}}\cdot|I_{\mathcal{S}}|\right),i\in I_{\mathcal{S}}\right\}
\end{align}

All remaining items are debiased perspective items, denoted as $I_{\mathcal{D}}^\mathcal{S}$:
\begin{equation}
    I_{\mathcal{D}}^{\mathcal{S}}=I_{\mathcal{S}}-I_{\mathcal{P}}^{\mathcal{S}}-I_{\mathcal{A}}^{\mathcal{S}}
\end{equation}

\begin{figure*}[ht]
    \centering
    \includegraphics[width=1\linewidth]{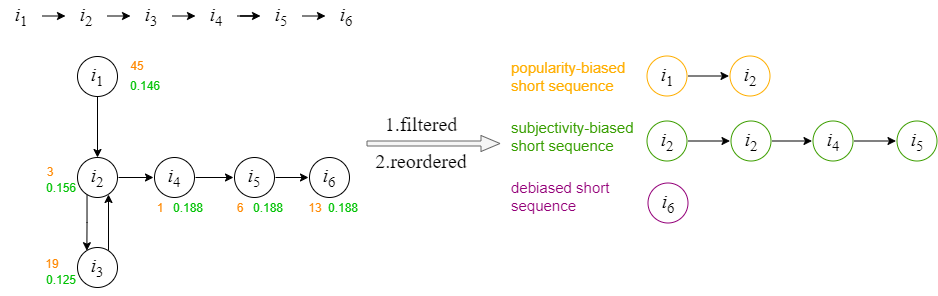}
    \caption{Short sequence recombination}
    \label{fig: recombination}
\end{figure*}

It is worth noting that the forward and backward order of the items in the sequence does not change after the items are filtered and reorganized. Figure \ref{fig: recombination} shows the process of screening and reorganization of sequence data where the values of $k_{\mathcal{P}}$, $k_{\mathcal{A}}$ are both 0.5. 

\subsection{Project Diagram Construction}
Due to the short length of the sequences, it is difficult to provide sufficient contextual signals for the neural sequence encoder, thus limiting the performance of the model. To solve this problem, we utilize the construction method of item graph structure. For three short sequences with different biases, the same construction method is used to obtain graph structures ((i.e., $\mathcal{G}_{\mathcal{P}}$, $\mathcal{G}_{\mathcal{A}}$, and $\mathcal{G}_{\mathcal{D}}$) under three different bias perspectives. This construction of graph structures can learn cross-sequence dependencies between individual users under the same bias, which can help the neural network to better understand the users' behavioral patterns under different bias perspectives. 

Specifically, the neighboring item pairs in each sequence $S$ are treated as edges of the item graph $\mathcal{G}$, and this view is represented using an adjacency matrix. The adjacency matrix $A_{\mathcal{G}}\in\mathbb{R}^{|I|\times|I|}$ is generated according to the following formula: 
\begin{equation}
    A_{\mathcal{G}}^u(i_p,i_q){=}\begin{cases}1,|p-q|=1\\0,otherwise&\end{cases};A_{\mathcal{G}}=\sum_{u\operatorname{=}1}^{|U|}A_{\mathcal{G}}^u
\end{equation}
where $A_{\mathcal{G}}^u$ denotes the neighboring item transitions of the specified user in the sequence $S_u$. $p$, $q$ denote the index position of the item $i$ in the sequence, and the corresponding adjacency matrix element is 1 if item $_p$ and item $i_q$ are adjacent in the sequence, and 0 otherwise. The complete adjacency matrix $A_{\mathcal{G}}$ is obtained by superimposing all the user's $A_{\mathcal{G}}^u$. It is important to emphasize that each value in the adjacency matrix $A_{\mathcal{G}}$ represents the number of transitions between neighboring items, and thus corresponds to the weights of the upper edges of the view $\mathcal{G}$. 

After the above steps, the adjacency matrices of the graphs in the three bias perspectives are obtained as $A_{\mathcal{G}_{p}}$,$A_{\mathcal{G}_{A}}$, and $A_{\mathcal{G}_{D}}$. To emphasize the importance of each project node, the unit diagonal matrices of the three adjacency matrices are summed up to obtain new adjacency matrices $\tilde{A}_{\mathcal{G}_{p}}$, $\tilde{A}_{\mathcal{G}_{A}}$, and $\tilde{A}_{\mathcal{G}_{p}}$. Then, the rows of the adjacency matrices are summed up to obtain the degree matrices $D_{\mathcal{G}_{p}}$,$D_{\mathcal{G}_{A}}$, and $D_{\mathcal{G}_{D}}$. The degree matrices reflect the degree of each node, i.e., the number of edges connected to it. 

\begin{figure*}[ht]
    \centering
    \includegraphics[width=0.5\linewidth]{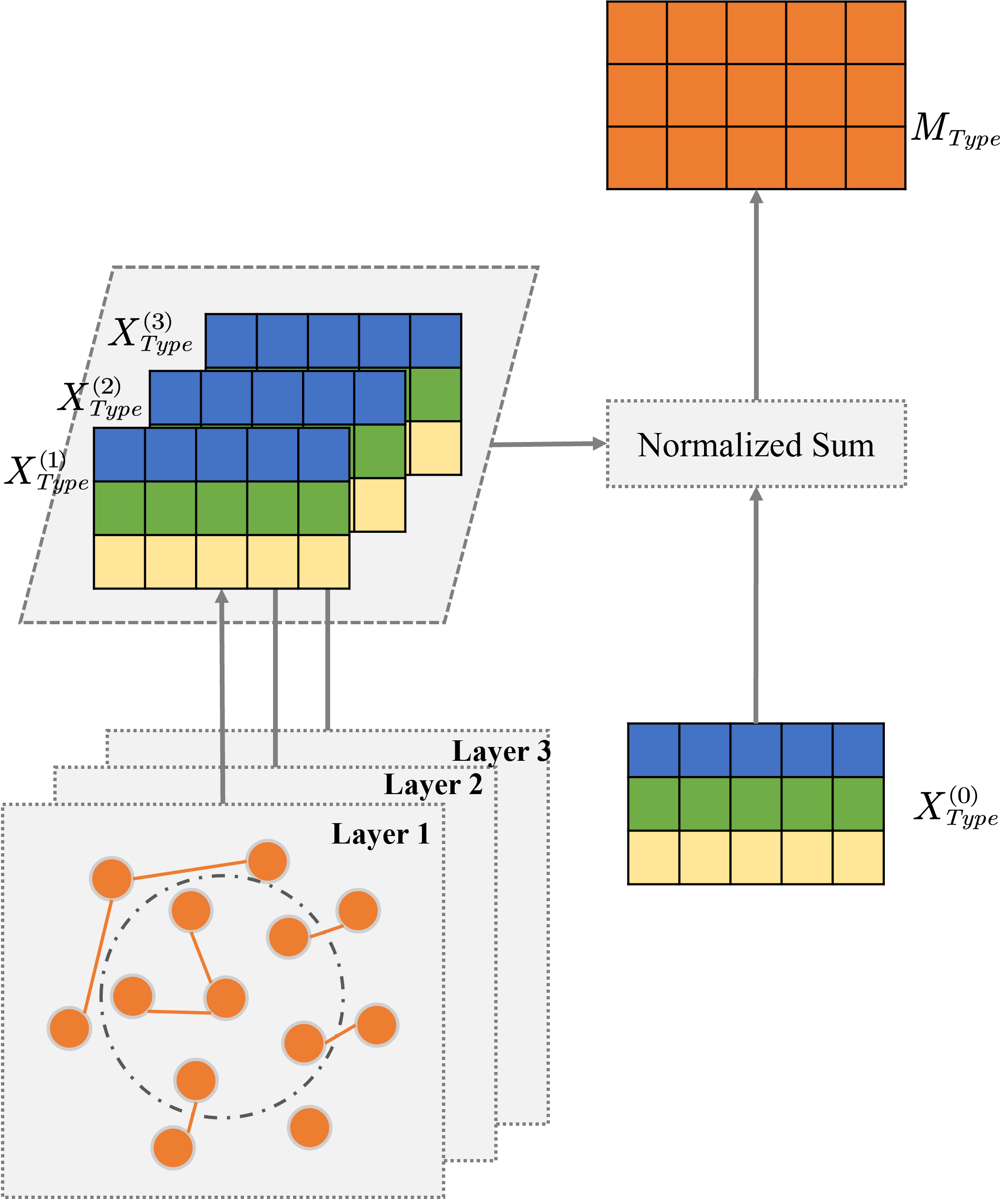}
    \caption{graph structure message passing aggregation }
    \label{fig: graph}
\end{figure*}
To simplify the model and improve its efficiency and interpretability, the model employs a strategy of culling the nonlinear activation operations and redundant transformation operations in the Graph Convolutional Network (GCN). 
The formula of the whole message-passing process is represented as follows:
\begin{equation}
    X_{Type}^{(l+1)}=\left(D_{Type}^{-1/2}A_{{\mathcal{G}_{Type}}}D_{Type}^{-1/2}\right)X_{Type}^{(l)}
\end{equation}
where $l$ denotes the level index, and $Type$ denotes the type of bias. $X_{Type}^{(l)}\in\mathbb{R}^{|I|\times d}$ denotes the feature matrix after $l$ layers of message passing under some bias perspective, where $d$ denotes the dimension size of each embedding vector. As shown in Figure \ref{fig: graph}, in the initial state, the feature matrix $X_{Type}^{(0)}$ consists of the embedding matrices of all items. Then, the embedding feature information constructed based on the graph can be obtained by averaging the feature matrices of each layer by superposition. The embedded feature information $M_{type}$ constructed based on the graph is calculated as follows: 
\begin{equation}
    M_{type}=\frac{\sum_{l=1}^{L}X_{Type}^{(l)}}{L}
\end{equation}
where $L$ denotes the number of total layers of message delivery.

\subsection{Sequence Information Encoding}
The three biased feature matrices degree (i.e., $A_{\mathcal{G}_{p}}$,$A_{\mathcal{G}_{A}}$, and $A_{\mathcal{G}_{D}}$) obtained after graph message-passing process contain the potential vector information of the items in three biased perspectives. To record the location information of items at different locations in the sequence, we introduce a learnable location matrix $P\in\mathbb{R}^{L\times d}$. Then the potential vector representation $E\in\mathbb{R}^{L\times d}$ after location information embedding is defined as follows:
\begin{equation}
    E=\begin{bmatrix}M_{s_1}+P_1\\M_{s_2}+P_2\\...\\M_{s_L}+P_L\end{bmatrix}
\end{equation}
where $M_{s_{i}}$ denotes the potential vector representation of the $i$-th item in the sequence $s$. In this way, the item representation at each position contains its potential features and the embedding of the positional information, which enables the model to better understand and utilize the information of the items at different positions in the sequence. 

\begin{figure*}[ht]
    \centering
    \includegraphics[width=1\linewidth]{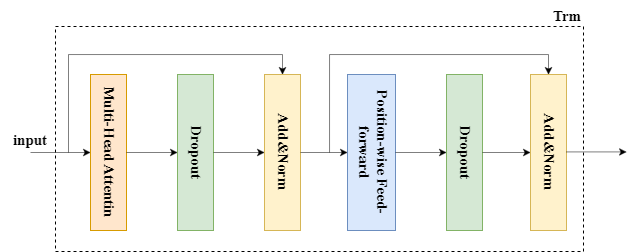}
    \caption{Transformer framework}
    \label{fig: transformer}
\end{figure*}

For the model to better understand and capture the interactions and dependencies between different items in a sequence of user behaviors, MABSRec introduces the Transformer module to encode the sequence information under multiple bias perspectives. Figure \ref{fig: transformer} illustrates the framework of the module.

Firstly, this module performs a Dropout operation on the obtained latent vector representation after location information embedding $E$ and the feature matrix after the Dropout operation is noted as $\tilde{E}$. 
After that, the multi-head self-attention mechanism is employed to capture the contextual information of the sequence, which operates as follows: 
\begin{equation}
     \text{MultiHead(H)}=\text{Concat}(head_{1},head_{2},\cdots,head_{h})\cdot W_{o}
\end{equation}
\begin{equation}  head_i=\text{Attention}\left(\tilde{E}W_{Q_i},\tilde{E}W_{K_i},\tilde{E}W_{V_i}\right)
\end{equation}
\begin{equation}
    \mathrm{Attention}(Q,K,V)=\mathrm{Dropout}\left(\mathrm{softmax}\left(\frac{QK^T}{\sqrt{d/h}}\right)\right)V
\end{equation}
where $W_Q\:\text{,}\:W_K\:\text{,}\:W_V\in\mathbb{R}^{d\times\frac dh}$ correspond to the head-specific mapping matrices for queries, keys, and values, respectively. $W_o$ is the parameter output matrix with dimension $d \times d$, and $h$ is the number of attention heads. In the multi-head self-attention mechanism, the linear transformations of the query, key, and value are first computed separately through $h$ heads. 
During the computation of the model, for the results after the foregoing transformations, the scaled dot product attention mechanism is used to perform further processing. 
Finally, the outputs of all attention heads are concatenated and linearly transformed through a weight matrix $W_o$ to obtain the final output $x\in\mathbb{R}^d$ of the multi-head self-attention mechanism. To enhance the gradient flow of the model and facilitate the learning of sequence representations during training, we apply normalization and residual connections to the output of the multi-head self-attention mechanism, denoting the result as $\tilde{x}$.
In addition, a feed-forward neural network is introduced to transform the representation to make the model learn the effect of non-linearity:

\begin{equation}
    \mathrm{FFN}(\tilde{x})=\mathrm{GELU}(\tilde{x}W_1+b_1)W_2+b_2
\end{equation}
where $W_1\text{,}W_2\in\mathbb{R}^{d\times d}\:\text{,}\:b_1\:\text{,}\:b_2\in\mathbb{R}^d$ are the weight and bias matrices, respectively. $\mathrm{GELU}(\cdot)$ is the nonlinear activation function. 

It is worth noting that for all three biased feature inputs, the model chooses the same Transformer module to receive them. This approach enables the model to share parameters when processing sequence information under different biases, reducing the model's complexity and the number of parameters. By sharing the Transformer module, the model can better learn the sequence features under different biases and generalize to new datasets, improving the model's generalization ability. 

\subsection{Adaptive Multi-bias Perspective Attention Module}
To better fuse the three biased sequence information for recommendation, we propose adaptive multi-perspective attention learning. Specifically, a vector addition operation is first adopted to fuse these bias features so that the model can not only capture the impact of each single bias but also reveal the possible mutual enhancement or mutual inhibition among different biases. As follows, we define the feature vectors that fuse the three bias information: 
\begin{align}
     o_{u}=\mathrm{Concat}[x_{\mathcal{P}_{*}},x_{\mathcal{A}_{*}},x_{\mathcal{D}_{*}},x_{\mathcal{P}_{*}}\oplus x_{\mathcal{A}_{*}},\\ \notag x_{\mathcal{P}_{*}}\oplus x_{\mathcal{D}_{*}},x_{\mathcal{D}_{*}}\oplus x_{\mathcal{D}_{*}}\oplus x_{\mathcal{D}_{*}}]
\end{align}where $\oplus$ denotes vector summation. Then, for each user sequence $s_u$ the corresponding feature vector $o_u$ will be fed into a two-layer feed-forward neural network, and then its long dimensional information will be nonlinearly processed by the ReLU function, which enables the model to better focus on the learning of key features. Next, the Sigmoid function is utilized to output the scores $scores_u$ corresponding to the feature information under the three bias perspectives. These scores can indicate the degree of user preference for items under different bias perspectives, thus providing critical information for the recommendation model. The user's score $scores_u$ for the three bias perspectives is calculated as follows:
\begin{equation}
    scores_{u}=\sigma(W_{2}\mathrm{ReLU}(W_{1}o_{u}+b_{1})+b_{2})
\end{equation}where $W_{1}\in\mathbb{R}^{{d_{o}\times d}}\text{,}W_{2}\in\mathbb{R}^{d\times3}\text{,}b_{1}\in\mathbb{R}^{d}\text{,}b_{2}\in\mathbb{R}^{3}$ are all learnable parameters shared across sequence terms, $\sigma $ refers to the Sigmoid activation function, and $o_u$ refers to the dimension of $d_o$. Finally, the vector representation of the predicted items $e_{u}^{pred}$ is obtained by matrix multiplication of $scores_u$ with the corresponding eigenvectors $x_{\mathcal{P}_u}\text{,}x_{\mathcal{A}_u}\text{,}x_{\mathcal{D}_u}$ under the three bias perspectives.
\begin{equation}
    e_{u}^{pred}=\mathrm{matmul}\left(scores_{u},[x_{{\mathcal{P}_{{_{u}}}}}\|x_{{\mathcal{A}_{{\mathcal{u}}}}}\|x_{{\mathcal{D}_{{_{u}}}}}]\right)
\end{equation}
where $\mathrm{matmul}$ denotes the matrix multiplication, such that the resulting $e_{u}^{pred}$ represents the predicted preference of user $u$ for items with different bias perspectives.

\subsection{Loss Function}
In the training phase, the model is trained by using the last item in the user's interaction sequence as a label. To predict the next interaction item of the user, a recommendation score vector $\hat{y}_{u,i}$ is obtained by transposing and multiplying the predicted item vector $e_{u}^{pred}$ with the initial embedding representation $E\in\mathbb{R}^{|I|\times d}$ of all items as follows:
\begin{equation}
    \hat{y}_{u,i}=e_u^{pred}\cdot E^T
\end{equation}

The vector $\hat{y}_{u,i}$ represents the user's preference for all items under the current bias perspective. For each user's true next interaction item $i_{t+1}$, we define the recommendation error $\mathcal{L}_{rec}$:
\begin{equation}
    \mathcal{L}_{rec}=-\log\left(\frac{\exp\left(\hat{y}_{{u,i_{t+1}}}\right)}{\sum_{j\in I}\exp\left(\hat{y}_{u,j}\right)}\right)
\label{loss}
\end{equation}

This loss function measures the extent to which the probability distribution of items predicted by the model deviates from the true interaction items through the cross-entropy loss function. In Equation (\ref{loss}), the negative log-likelihood loss will be used to minimize the gap between the predicted probabilities and the true labels, where the numerator part $\exp(\hat{y}_{u,i_{t+1}})$ denotes the predicted probability of the next item, and the denominator part is the softmax normalization of the probabilities of all items. Optimizing the model parameters by minimizing the recommendation error loss makes it more accurate in predicting the user's behavior. 

\section{Experimental Results and Analysis}
\label{experimental}
In this section, we evaluate our algorithm comprehensively through experiments.
\subsection{Experiment Settings}
\subsubsection{Dataset}
To validate the effectiveness of the model, we evaluated it on three real-world benchmark datasets, i.e., Amazon Beauty, Amazon Sports, and MovieLens-20M datasets. The statistics for various datasets are provided in Table \ref{tab: dataset}. 
\begin{itemize}
    \item The \textbf{Amazon Beauty} dataset is obtained from the Amazon website and contains detailed behavioral records and product characteristic information of users when purchasing beauty products.
    \item The \textbf{Amazon Sports} dataset is a dataset for the Amazon Outdoor Sports product segment, similar to the Amazon Beauty dataset, which also contains user purchase and review information.
    \item The \textbf{MovieLens-20M} dataset is a classic movie rating dataset that contains rating information for over 12,000 movies from over 50,000 users.
\end{itemize}   
To ensure the reliability of the data, interactions with sequence lengths less than 5 were eliminated from the experiments. Also for the MovieLens-20M dataset, interactions with sequence lengths more than 50 were removed. 
\begin{table}[t]
    \centering
    \caption{Statistics of the Datasets}
    \resizebox{1\linewidth}{!}{
    \begin{tabular}{c|c|c|c|c|c} \cline{1-6}
         Datasets   &\#users   &\#items   &\#Interactions   &\#Avg.leng &\#Density  \\ \cline{1-6}
         Beauty   &52374   &121290   &469771  &8.97 &7.4e-5\\ \cline{1-6}
         Sports  &84368  &194714  &717464  &8.50 &4.4e-5 \\ \cline{1-6}
         ML-20M  &54437  &12360   &1727055   &31.73  &2.6e-3 \\ \cline{1-6}
    \end{tabular}
    }
    \label{tab: dataset}
\end{table}

\subsubsection{Baselines}
To validate the effectiveness of our model, we conducted comparative experiments with several classical baseline recommendation models and excellent recommendation models that have recently received much attention. The compared baselines are described as follows:
\begin{itemize}
    \item \textbf{Caser \cite{10.1145/3159652.3159656}:} This is a convolutional neural network-based recommendation model that focuses on learning local feature information of sequence data. 
    \item \textbf{GRU4Rec: \cite{hidasi2016sessionbasedrecommendationsrecurrentneural}} A recommender system based on session modeling techniques, the core idea of which is to deeply learn the user's behavioral sequences through a multilayered GRU layer and a fully-connected layer to capture the temporal dynamics of the user's behavior and the potential interest evolution.
    \item \textbf{SASRec: \cite{8594844}} This model introduces the Transformer model to the field of recommender systems, mainly by introducing positional embedding and self-attention mechanisms to deal with sequential data.
    \item \textbf{CL4SRec \cite{9835621}:} It innovatively performs censoring, masking, and disruption operations on the items of user sequences to generate enhanced data. 
    \item \textbf{DuoRec \cite{10.1145/3488560.3498433}:} It utilizes different Dropout techniques to augment the data while training sequences with the same target as positive samples. 
    \item \textbf{MAERec \cite{10.1145/3539618.3591692}:} Unlike traditional randomly augmented data, MAERec can dynamically select nodes to be masked based on the connectivity and importance of the nodes in the graph, thus generating more representative training samples. 
    \item \textbf{DCRec \cite{10.1145/3543507.3583361}:} Build a cross-view comparative learning framework that aims to learn the degree of user follower and normalize the follower distribution of all users to a normal distribution through KL dispersion to achieve the effect of debiasing.
\end{itemize}

\subsubsection{Evaluation Indicators}
During the model evaluation process, this chapter uses a Top-N correlation recommendation list and employs two main evaluation metrics: Recall Rate @N (Recall@N)  and Normalized Discounted Cumulative Gain @N (NDCG@N), where N $\in \{1, 5, 10\}$. 

Recall@N measures the model's ability to capture a user's true interest given a recommendation list of length N. The Recall@N of the entire dataset is the average of all users' Recall@N, providing a global view of the model's performance. Recall@N is calculated as follows:

NDCG@N considers the positional factors of the items in the recommendation list and measures the quality of the model's ranking in the recommendation list by calculating the Discounted Cumulative Gain (DCG). 


\subsection{Parameter Settings}
The batch size of all datasets is set as 512 and the sequence length is fixed to 50. To ensure the fairness of the experiments, an Adam optimizer with a learning rate of 0.001 is used for all the training models and an early-stopping strategy is set up: when the model fails to show any better metrics within 10 epochs, the training is stopped. In the MABSRec model, the parameters are set differently for different dataset characteristics. On both the Amazon Beauty and Amazon Sports datasets, the dropout rate is 0.4, and the number of Transformer layers and graph convolution layers is set as 2. On the MovieLens-20m dataset, due to the different characteristics of the dataset, the dropout rate is set as 0.1, and the number of Transformer layers and graph convolution layers is increased to 4. In addition, the number of multi-head self-attention heads is set as 1 on both the Amazon Beauty and Amazon Sports datasets, while it is increased to 8 on the MovieLens-20M dataset. The graph convolution dropout rates are 0.4, 0.5, and 0.3, respectively. Table \ref{tab: parameter} demonstrates the parameter settings in this experiment.


\begin{table}[t]
    \centering
    \caption{PARAMETER SETTINGS}
    \resizebox{1\linewidth}{!}{
    \begin{tabular}{c|c|c|c} \cline{1-4}
        ~ &Beauty  &Sports  &ML-20M \\ \cline{1-4}
         batch size&512   &512   &512  \\ \cline{1-4}
         Adam learning rate&0.001   &0.001   &0.001  \\ \cline{1-4}
         maximum sequence length&50   &50   &50  \\ \cline{1-4}
         dropout rate&0.4  &0.4  &0.1 \\ \cline{1-4}
         number of Transformer layers&2   &2   &4 \\ \cline{1-4}
         \makecell[c]{number of multi-head \\self-attention heads}&1  & 1 & 8\\ \cline{1-4}
         number of graph convolution layers& 2 & 2& 4\\ \cline{1-4}
         graph convolution dropout rate&0.4  & 0.5 &0.3 \\ \cline{1-4}
    \end{tabular}
    }
    \label{tab: parameter}
\end{table}

\subsection{Comparative Experiment}
In this section, we conduct comparative experiments of the proposed method with the baseline model and select relevant evaluation metrics for assessment.

\subsubsection{Performance Comparison}
For the performance of various models, we experimented MABSRec with other baseline models on different datasets and evaluated them according to different evaluation metrics (Recall@1, Recall@5, Recall@10, NDCG@5, NDCG@10). The experiment results are shown in table \ref{tab: compared}.
\begin{table*}[ht]
    \centering
    \caption{Experimental results for each model on three real datasets }
    \resizebox{1\linewidth}{!}{
    \begin{tabular}{c|c|c|c|c|c|c|c|c|c|c} \hline
         \makecell{Dataset}&Metric  &Caser   &GRU4Rec   &SASRec   &CL4SRec   &DuoRec   &MAERec   &DCRec  &\textbf{MABSRec}  &\#improve \\ \hline
         \multirow{5}{*}{\begin{tabular}[c]{@{}l@{}}Beauty\end{tabular}}
         &Recall@1   &0.0098  &0.0101  &0.0107  &0.0145  &\underline{0.0147}  &0.0138  &0.0118  &\textbf{0.0149}  &\textbf{1.36\%} \\
         &Recall@5   &0.0226  &0.0240  &0.0251  &0.0326  &\underline{0.0328}  &0.0327  &0.0272  &\textbf{0.0336}  &\textbf{2.44\%}\\
         &Recall@10   &0.0336  &0.0341  &0.0377  &0.0437  &0.0441  &\underline{0.0445}  &0.0362  &\textbf{0.0453}  &\textbf{1.80\%}\\
         &NDCG@5  &0.0175  &0.0183  &0.0202  &0.0237  &\underline{0.0238} &0.0232  &0.0196  &\textbf{0.0242}  &\textbf{1.68\%}\\
         &NDCG@10  &0.0209  &0.0210  &0.0234  &0.0272  &0.0275    &0.0272  &0.0225  &\textbf{0.0280}  &\textbf{1.82\%}\\ \hline
         \multirow{5}{*}{\begin{tabular}[c]{@{}l@{}}Sports\end{tabular}}
         &Recall@1   &0.0045  &0.0046  &0.0050  &0.0065  &\textbf{\underline{0.0071}}  &0.0060  &0.0053  &\textbf{0.0071}  &\textbf{0.00}\%\\
         &Recall@5   &0.0117  &0.0120  &0.0129  &\underline{0.0168}  &0.0165  &0.0163  &0.0129  &\textbf{0.0176} &\textbf{4.76\%} \\
         &Recall@10   &0.0163  &0.0169  &0.0186  &\underline{0.0243}  &0.0237  &0.0236  &0.0183  &\textbf{0.0251}  &\textbf{3.29\%} \\
         &NDCG@5  &0.0083  &0.0086  &0.0090  &0.0117  &\underline{0.0118}  &0.0112  &0.0091  &\textbf{0.0123}  &\textbf{4.24\%} \\
         &NDCG@10  &0.0098  &0.0101  &0.0108  &0.0141  &\underline{0.0142}  &0.0135  &0.0109  &\textbf{0.0147}  &\textbf{3.52\%} \\ \hline
         \multirow{5}{*}{\begin{tabular}[c]{@{}l@{}}ML-20M\end{tabular}}
         &Recall@1   &0.0908   &0.0899   &0.0954   &0.1032   &0.1031  &0.0673   &\underline{0.1064}  & \textbf{0.1100}  &\textbf{3.38\%} \\
         &Recall@5   &0.2182   &0.2134  & 0.2259 &0.2364  &0.2367   & 0.1816 & \underline{0.2415}  & \textbf{0.2463}  &\textbf{1.99\%} \\
         &Recall@10   &0.2974  & 0.2875  &0.3053   &0.3203   &0.3204   &0.2589  &\underline{0.3224}   &\textbf{0.3289}  &\textbf{2.02\%} \\
         &NDCG@5  &0.1532   &0.1488  &0.1624   & 0.1716  & 0.1719  & 0.1257 &\underline{0.1759}  & \textbf{0.1798} &\textbf{2.22\%} \\
         &NDCG@10  & 0.1825  &0.1797   &0.1880   &0.1986   &0.1989   &0.1506  & \underline{0.2020} &\textbf{0.2065}  &\textbf{2.23\%} 
\\ \hline
    \end{tabular}
    }
    \label{tab: compared}
\end{table*}

Based on the comparison results shown in table \ref{tab: compared}, the recommendation models with attention mechanisms significantly outperform traditional models based on RNN or CNN. Therefore, we focus on two evaluation metrics, Recall@10 and NDCG@10, to conduct in-depth performance analysis and exploration of recommendation models (e.g., SASRec, CL4sRec, Duorec, MAERec, DCRec, and MABSRec) that incorporate the attention mechanism, and to further reveal the strengths and characteristics of these models in recommender systems.

\begin{figure*}[htbp]
	\centering
	\begin{minipage}{0.32\linewidth}
		\centering
		\includegraphics[width=0.9\linewidth]{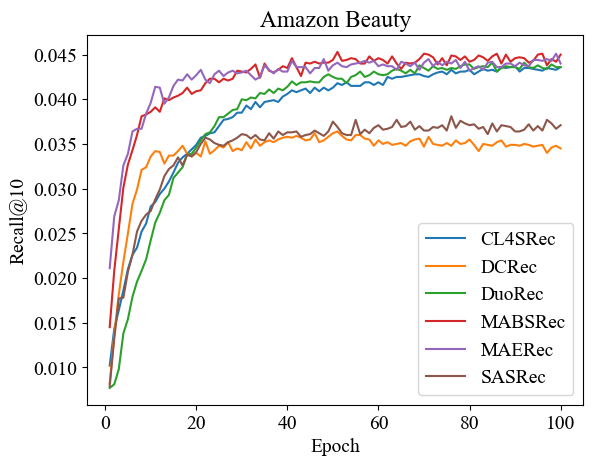}
	\end{minipage}
	\begin{minipage}{0.32\linewidth}
		\centering
		\includegraphics[width=0.9\linewidth]{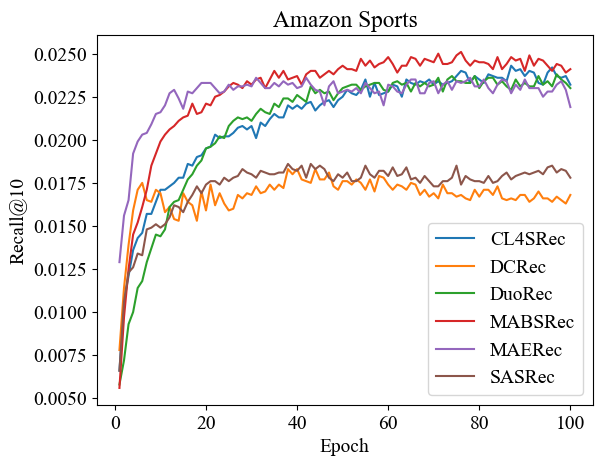}
	\end{minipage}
    \begin{minipage}{0.32\linewidth}
		\centering
		\includegraphics[width=0.9\linewidth]{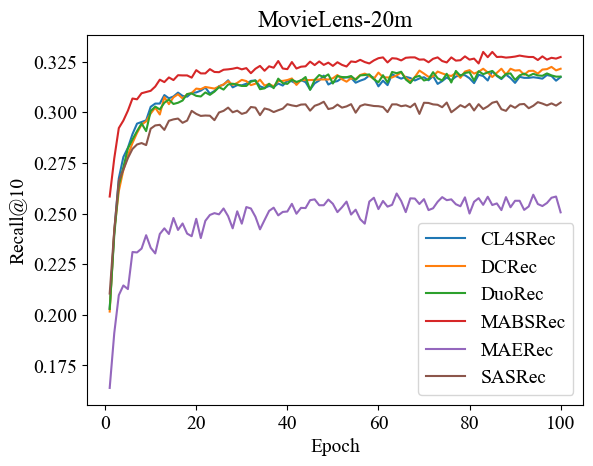}
	\end{minipage}
    

    \begin{minipage}{0.32\linewidth}
		\centering
		\includegraphics[width=0.9\linewidth]{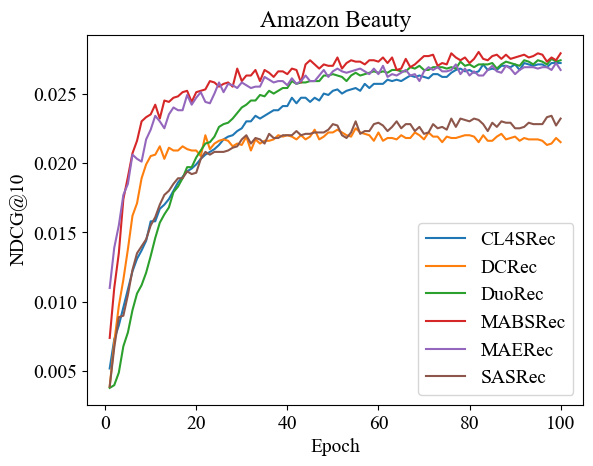}
	\end{minipage}
	\begin{minipage}{0.32\linewidth}
		\centering
		\includegraphics[width=0.9\linewidth]{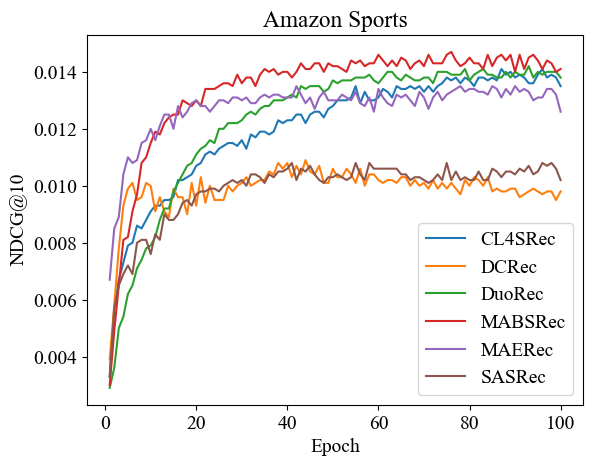}
	\end{minipage}
    \begin{minipage}{0.32\linewidth}
		\centering
		\includegraphics[width=0.9\linewidth]{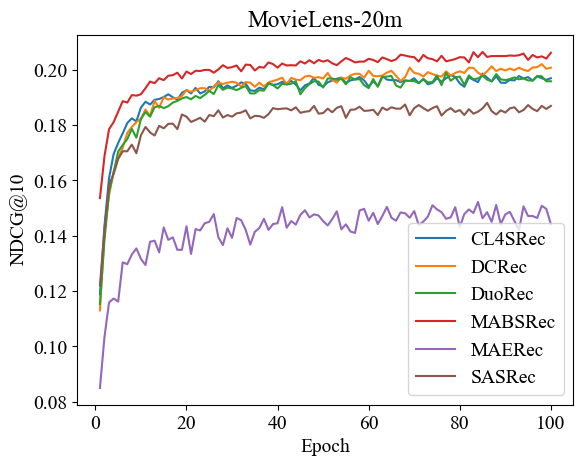}
	\end{minipage}
    \caption{Performance Trends on Three Datasets}
    \label{fig: epoch}
\end{figure*}

Figure \ref{fig: epoch} 
presents the training progress of the six recommendation models with attention modules on three different datasets. Overall, the MABSRec model shows excellent performance at the early stage of training on each dataset, and it can achieve high results after a relatively small number of Epochs, demonstrating an efficient training speed. As the training progresses, the performance of the MABSRec model continues to improve and shows high stability in the later stages of training, which is not prone to overfitting or performance degradation. This indicates that the model can fully mine the feature information of sequence data and effectively learn the user's preferences and interests. Therefore, from the comprehensive performance evaluation, the MABSRec model has important application prospects and research value in the field of recommender systems.

\begin{figure*}[htbp]
	\centering
    \begin{minipage}{0.32\linewidth}
		\centering
		\includegraphics[width=0.9\linewidth]{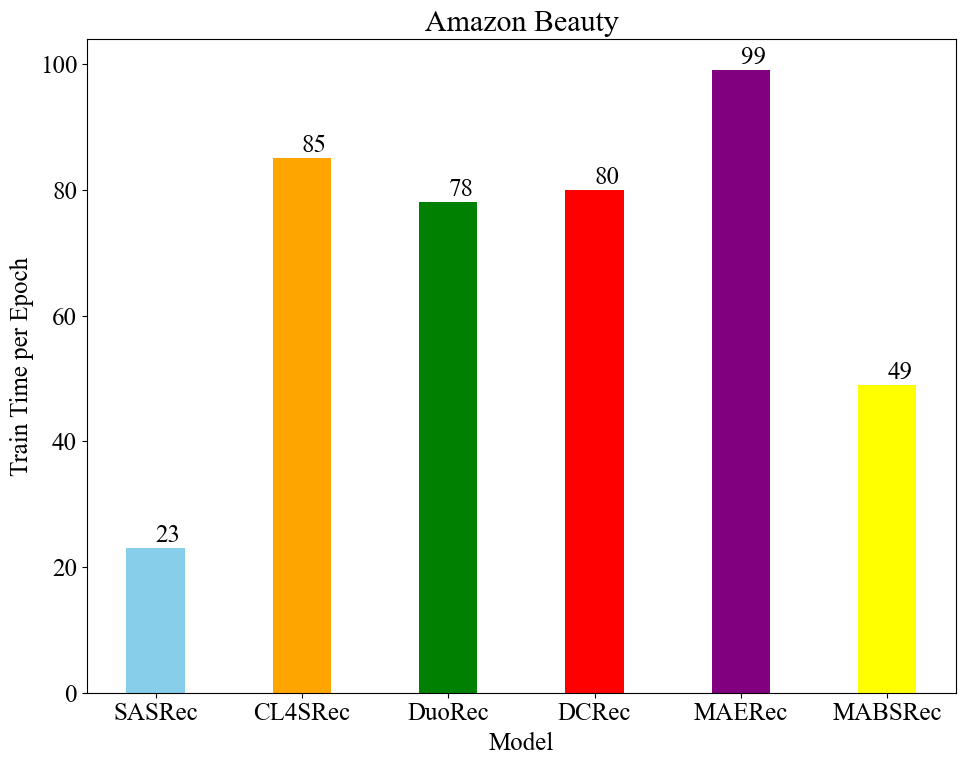}
	\end{minipage}
	\begin{minipage}{0.32\linewidth}
		\centering
		\includegraphics[width=0.9\linewidth]{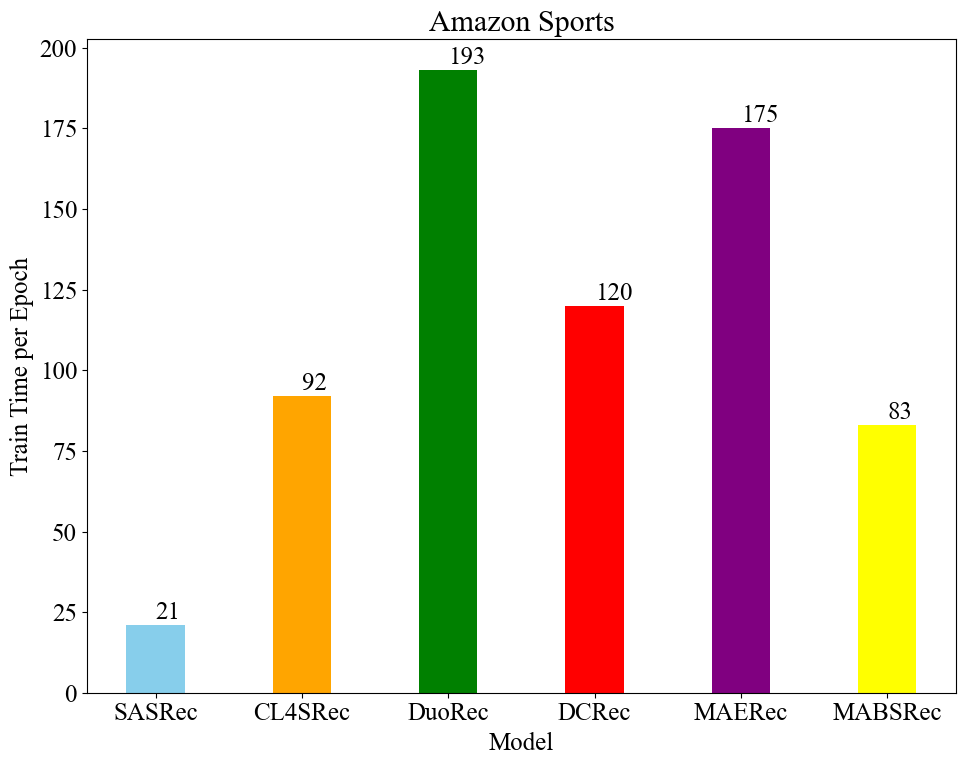}
	\end{minipage}
    \begin{minipage}{0.32\linewidth}
		\centering
		\includegraphics[width=0.9\linewidth]{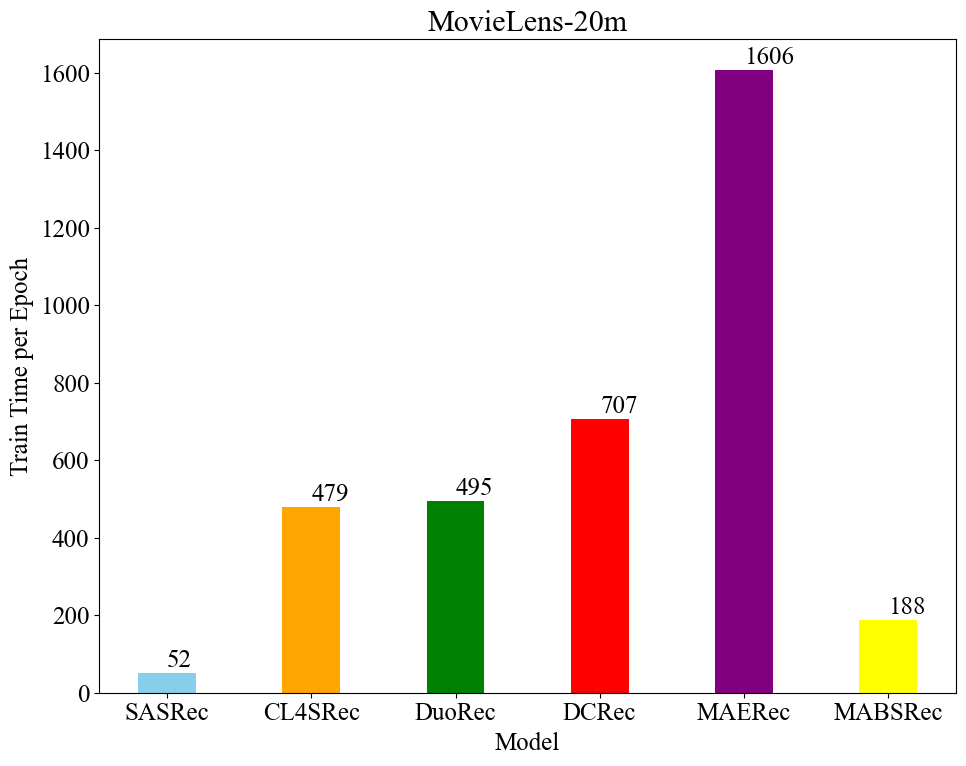}
	\end{minipage}
    \caption{Performance Trends on Three Datasets}
    \label{fig: time}
\end{figure*}

To more accurately assess the performance of each model, we compare the training time required to train an epoch on each dataset. The results presented in Figures \ref{fig: time} demonstrate a significant difference in the training time required by different recommendation models within a single Epoch. Notably, the SASRec model performs the best in terms of training time. This phenomenon is mainly attributed to the simplicity of the SASRec model, which does not involve complex comparison learning or graph structure processing operations. In contrast, the MABSRec model has a slightly longer training time. Although it introduces graph structure information, its main time consumption is still concentrated in the Transformer model part because it does not need to perform time-consuming operations such as graph sampling or random wandering. So compared to other graph structure-related recommendation models, the MABSRec model still shows relatively efficient training speed.

\subsubsection{Performance in Different Sequence Lengths}
Sequence length is an important parameter in recommender systems, which reflects the length of the interaction sequence between users and items. Sequences of different lengths may have an impact on the performance of the model. Therefore, this section investigates the performance of the MABSRec model with other recommendation models that incorporate the attention mechanism under different sequence length conditions. 

\begin{figure*}[ht]
	\centering
	\begin{minipage}{0.32\linewidth}
		\centering
		\includegraphics[width=0.9\linewidth]{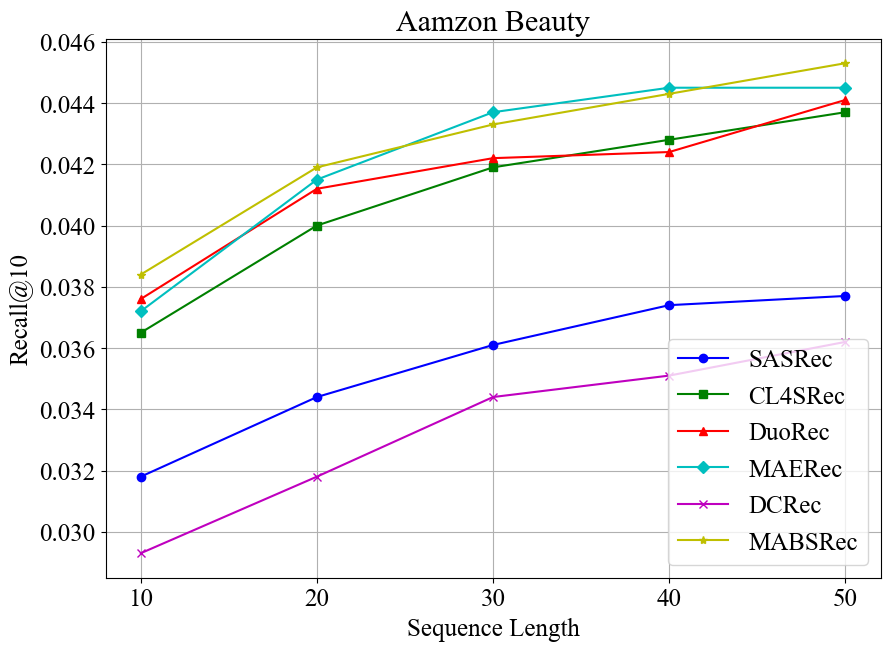}
	\end{minipage}
	\begin{minipage}{0.32\linewidth}
		\centering
		\includegraphics[width=0.9\linewidth]{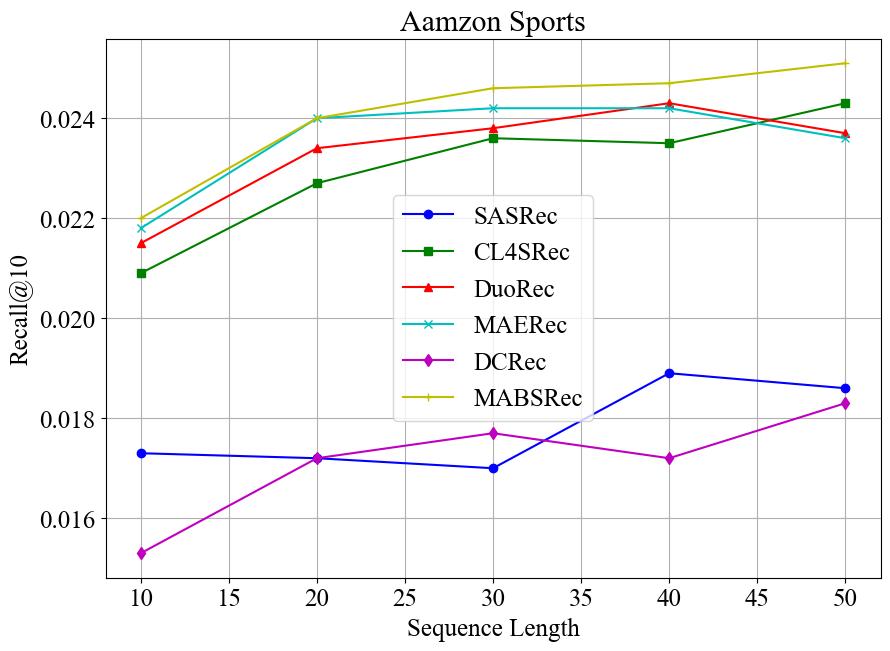}
	\end{minipage}
    \begin{minipage}{0.32\linewidth}
		\centering
		\includegraphics[width=0.79\linewidth]{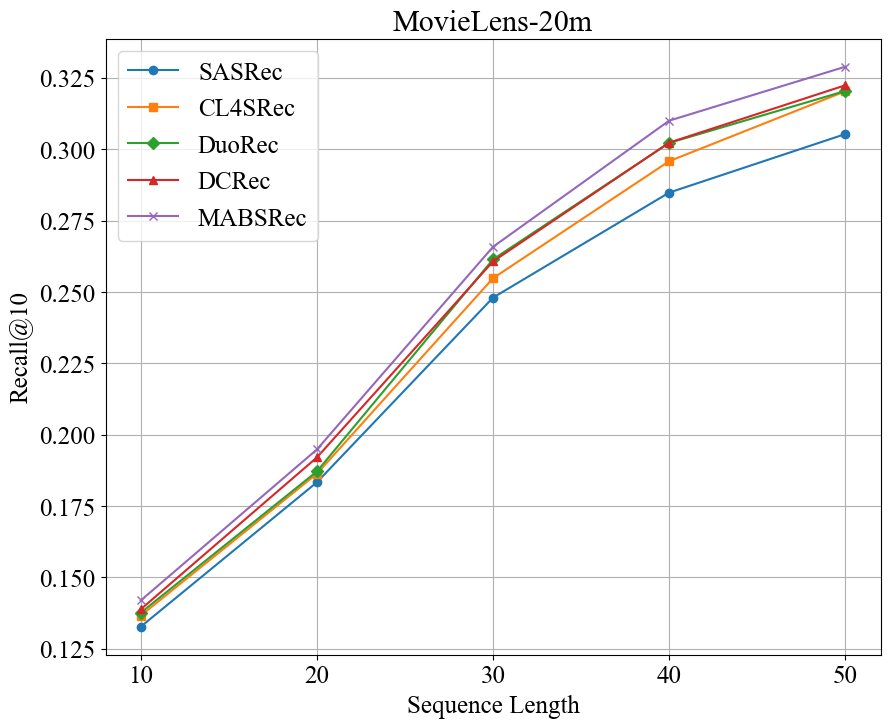}
	\end{minipage}
    

    \begin{minipage}{0.32\linewidth}
		\centering
		\includegraphics[width=0.9\linewidth]{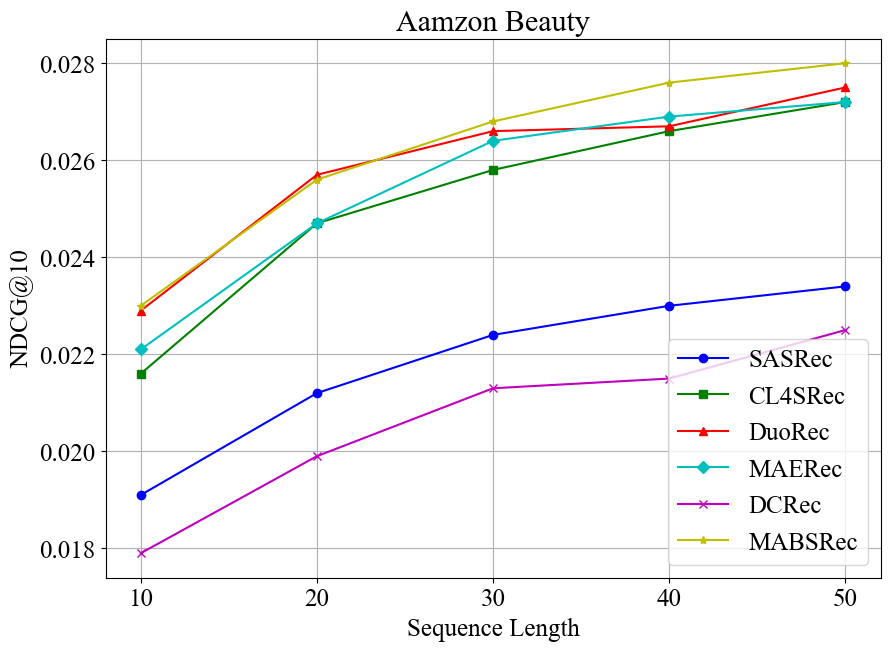}
	\end{minipage}
	\begin{minipage}{0.32\linewidth}
		\centering
		\includegraphics[width=0.9\linewidth]{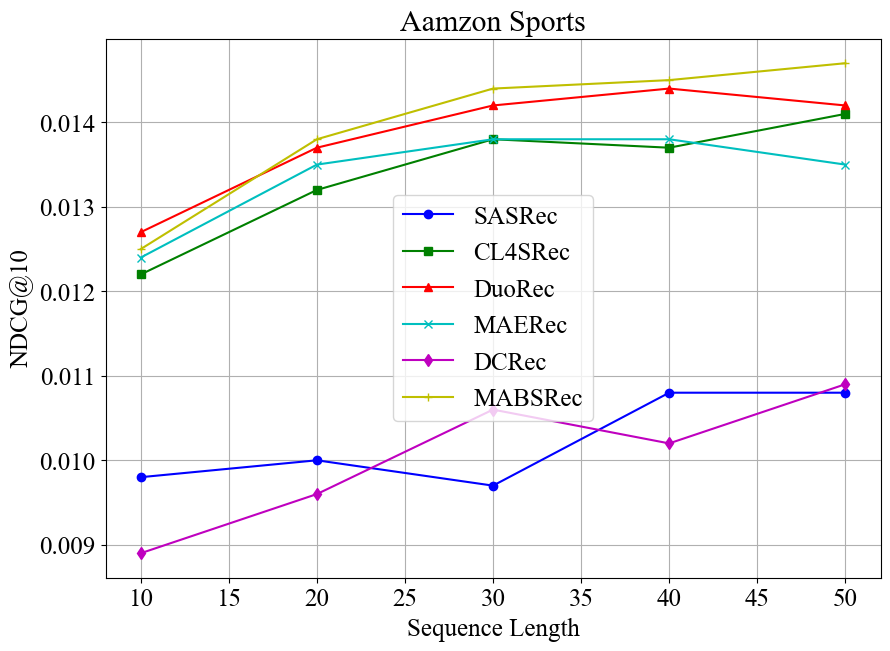}
	\end{minipage}
    \begin{minipage}{0.32\linewidth}
		\centering
		\includegraphics[width=0.79\linewidth]{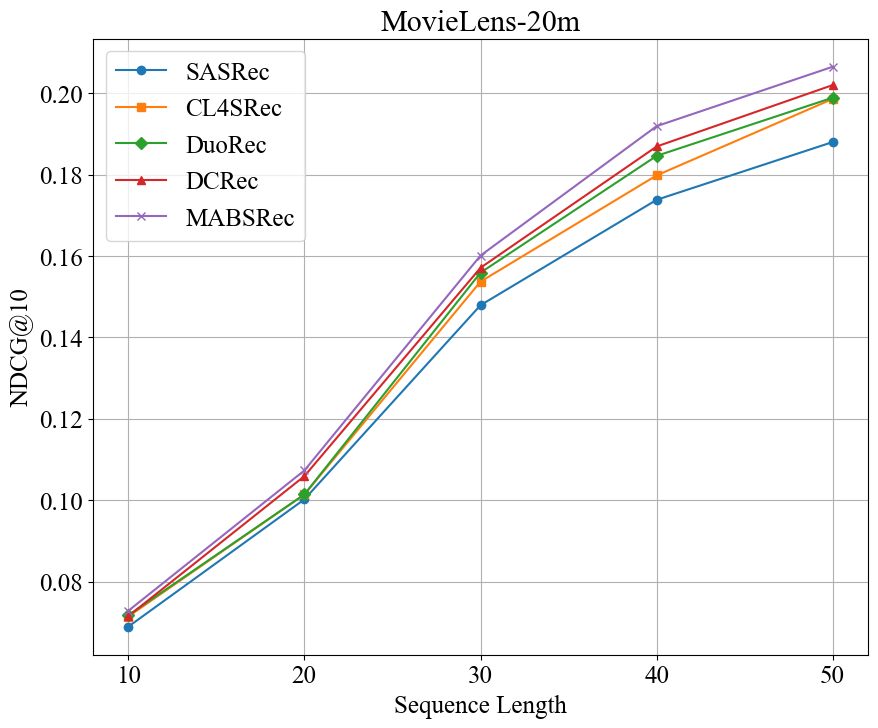}
	\end{minipage}
    \caption{Performance of Different Sequence Lengths}
    \label{fig: lengths}
\end{figure*}

Figure \ref{fig: lengths}  
shows the performance comparison graphs of each model under different datasets with different sequence lengths. The experimental results show that as the sequence length increases, the MABSRec model improves in various metrics and all of them show excellent performance, which is clearly ahead of all other models. This finding reflects the strong adaptability of the MABSRec model and shows that it is able to effectively capture user interests and behavioral patterns when dealing with different sequence lengths.

 
 

\subsubsection{Comparative Analysis of Bias Perspectives}
\begin{figure}[ht]
    \centering
    \includegraphics[width=1\linewidth]{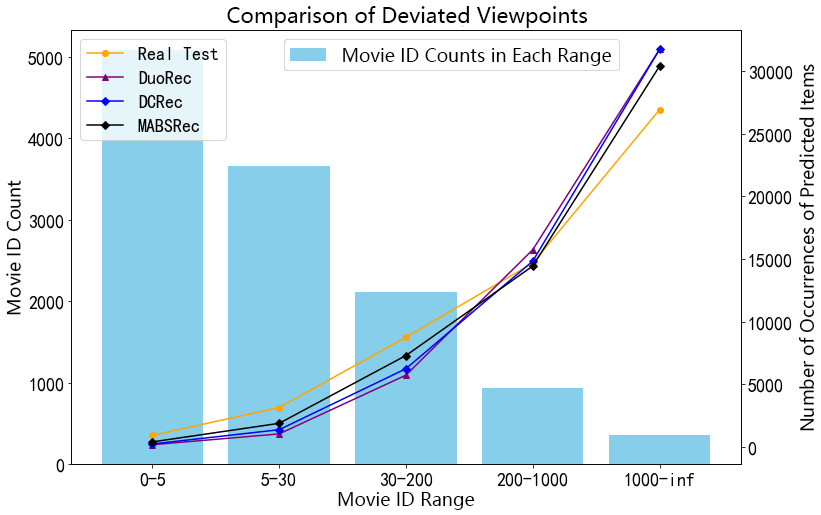}
    \caption{Comparison of Bias Perspectives}
    \label{fig: bias}
\end{figure}

To more clearly assess the necessity of multi-bias perspectives in the sequence recommendation problem, we compare MABSRec with the DuoRec and DCRec models that perform best on the MovieLens-20m dataset. The main reason for choosing these two models for comparison is that DuoRec is the best-performing model that does not consider the effect of any bias on the data, while DCRec fully considers the effect of popularity bias on the users. The experiment result is shown in Figure \ref{fig: bias}.
From the experimental results, compared with the DuoRec and DCRec models, the folds of the prediction results of the MABSRec model are significantly closer to the folds of the real test data. This indicates that the MABSRec model has higher accuracy and closeness in predicting item clicking behavior. This finding emphasizes the importance of multi-bias perspectives for recommender systems and provides useful guidance for further exploration of recommender models.

\subsubsection{Ablation Study}
To investigate the effect of key components on the model performance, ablation studies were performed on three variants of the MABSRec model: \textbf{w/o G}, \textbf{w/o A}, and \textbf{w/o D}. The results are presented in Table \ref{tab: ablation}.

\begin{table*}[ht]
    \centering
    \caption{Results of ablation experiments}
    \resizebox{1\linewidth}{!}{
    \begin{tabular}{ccccccc} \hline
         \multirow{2}{0.07\textwidth}{\centering Ablation Setting}&\multicolumn{2}{c}{Beauty}  &\multicolumn{2}{c}{Sports}  &\multicolumn{2}{c}{ML-20M}\\ \cline{2-7}
         &NDCG@5  &NDCG@10  &NDCG@5  &NDCG@10  &NDCG@5  &NDCG@10\\ \hline
         MABSRec  &0.0242  &0.0280  &0.0123  &0.0147  &0.1798  &0.2065\\ \hline
         w/o G  &0.0173 &0.0208 &0.0087 &0.0107 &0.1366 &0.1624\\ \hline
         w/o A  &0.0235 &0.0274 &0.0117 &0.0138 &0.1758 &0.2021\\ \hline
         w/o D  &0.0164 &0.0198 &0.0083 &0.0105 &0.1272 &0.1518\\ \hline
    \end{tabular}
    } 
    \label{tab: ablation}
\end{table*}

\textbf{Remove Graph Information (w/o G):} The first variant, \textbf{w/o G}, removes the graph information from the model consisting of biased short sequences. It trains the model directly using three short sequences as inputs to the model. The results show that graph information is crucial for capturing important structures and features in the sequences, and its absence leads to impaired model performance in the recommendation task.

\textbf{Remove Adaptive Multi-bias Perspective Attention Module (w/o A):} The Second variant, \textbf{w/o A}, replaces the process of learning information about the adaptive multi-bias perspective attention module with a simple average pooling operation. The results indicate that The adaptive multi-bias perspective attention module can effectively handle multiple types of bias information and improve the model's performance.

\textbf{Remove Both of the Above (w/o D):} The third variant, \textbf{w/o D}, removes both the graph information as well as the adaptive multi-bias perspective attention module from the model. The results show that the model performance degradation with these two components removed is more significant. This further validates the importance of the graph information and adaptive multi-bias perspective attention module on model performance.

\section{Conclusion}
\label{conclusion}
Current research on sequential recommender systems has received much attention. However, there is a lack of research on the impact of multi-bias factors in user data. To address this problem, we propose a novel sequence recommendation model, called MABSRec. The model first considers the prevalent popular and amplified subjective bias in user data and constructs a multi-bias view. Afterward, the degree of influence of various biases on the user is weighed by an attention fusion network to deeply mine the information in the user data and improve the overall performance of the recommender system. We validate the excellent performance of the MABSRec model by analyzing its metric scores on three real-world datasets compared to those of some excellent sequential recommendation models. The model proposed in this work can provide new ideas and methods for solving the biased problem in sequential recommender systems and also provides useful insights and directions for subsequent research on recommender systems.





\bibliographystyle{IEEEtran}
\bibliography{main}

\end{document}